\documentclass[twocolumn,trackchanges]{aastex62}

\hypersetup{citecolor=blue,filecolor=cyan,urlcolor=magenta,linkcolor=red}

\newcommand{\halpha}{H$\alpha$ }
\usepackage{amsmath}
\usepackage{multirow}
\usepackage[shortlabels]{enumitem}
\usepackage{natbib}

\begin{document}

\title{A COMPREHENSIVE STUDY OF H$\alpha$ EMITTERS AT $z \sim$ 0.62 IN THE DAWN SURVEY: \\THE NEED FOR DEEP AND WIDE REGIONS}

\author{Santosh Harish}
\affil{School of Earth and Space Exploration, Arizona State University, Tempe, AZ 85287, USA}
\email{santosh.harish@asu.edu}
\author{Alicia Coughlin}
\affil{Chandler-Gilbert Community College, 2626 East Pecos Road, Chandler, AZ 85225-2499, USA}
\author{James E. Rhoads}
\affil{School of Earth and Space Exploration, Arizona State University, Tempe, AZ 85287, USA}
\affil{NASA Goddard Space Flight Center, 8800 Greenbelt Road, Greenbelt, MD 20771, USA}
\author{Sangeeta Malhotra}
\affil{School of Earth and Space Exploration, Arizona State University, Tempe, AZ 85287, USA}
\affil{NASA Goddard Space Flight Center, 8800 Greenbelt Road, Greenbelt, MD 20771, USA}
\author{Steven L. Finkelstein}
\affil{Department of Astronomy, University of Texas at Austin, Austin, TX 78712, USA}
\author{Matthew Stevans}
\affil{Department of Astronomy, University of Texas at Austin, Austin, TX 78712, USA}
\author{Vithal S. Tilvi}
\affil{School of Earth and Space Exploration, Arizona State University, Tempe, AZ 85287, USA}
\author{Ali Ahmad Khostovan}
\affil{NASA Goddard Space Flight Center, 8800 Greenbelt Road, Greenbelt, MD 20771, USA}
\author{Sylvain Veilleux}
\affil{Department of Astronomy and Joint Space-Science Institute, University of Maryland, College Park, MD 20742 USA}
\affil{Institute of Astronomy and Kavli Institute for Cosmology Cambridge, University of Cambridge, Cambridge CB3 0HA, United Kingdom}
\author{Junxian Wang}
\affil{CAS Key Laboratory for Research in Galaxies and Cosmology, Department of Astronomy, University of Science and Technology of China, People's Republic of China}
\author{Pascale Hibon}
\affil{European Southern Observatory, Alonso de Cordova 3107, Vitacura, Casilla 19001, Santiago, Chile}
\author{Johnnes Zabl}
\affil{Univ Lyon, Univ Lyon1, Ens de Lyon, CNRS, Centre de Recherche Astrophysique de Lyon UMR5574, F-69230 Saint-Genis-Laval, France}
\author{Bhavin Joshi}
\affil{School of Earth and Space Exploration, Arizona State University, Tempe, AZ 85287, USA}
\author{John Pharo}
\affil{School of Earth and Space Exploration, Arizona State University, Tempe, AZ 85287, USA}
\author{Isak Wold}
\affil{NASA Goddard Space Flight Center, 8800 Greenbelt Road, Greenbelt, MD 20771, USA}
\author{Lucia A. Perez}
\affil{School of Earth and Space Exploration, Arizona State University, Tempe, AZ 85287, USA}
\author{Zhen-Ya Zheng}
\affil{Key Laboratory for Research in Galaxies and Cosmology, Shanghai Astronomical Observatory, Chinese Academy of Sciences, 80 Nandan Road, Shanghai 200030, People's Republic of China}
\author{Ronald Probst}
\affil{NOAO, 950 N. Cherry Avenue, Tucson, AZ 85719, USA}
\author{Rob Swaters}
\affil{Department of Astronomy, University of Maryland, College Park, MD 20742, USA}
\affil{Space Telescope Science Institute, 3700 San Martin Drive, Baltimore MD 21218, USA}
\author{Bahram Mobasher}
\affil{Department of Physics \& Astronomy, University of California, Riverside, USA}
\author{Tianxing Jiang}
\affil{School of Earth and Space Exploration, Arizona State University, Tempe, AZ 85287, USA}
\author{Huan Yang}
\affil{Las Campanas Observatory, Carnegie Institution for Science, Chile}




\begin{abstract}
We present new estimates of the luminosity function (LF) and star formation rate density (SFRD) for an \halpha selected sample at $z\sim0.62$ from the Deep And Wide Narrow-band (DAWN) survey. Our results are based on a new \halpha sample in the extended COSMOS region (compared to \citealt{Coughlin2018}) with the inclusion of flanking fields, resulting in a total area coverage of $\sim$1.5 deg$^2$. A total of 241 \halpha emitters were selected based on robust selection criteria using spectro-photometric redshifts and broadband color-color classification. Given that dust extinction is a dominant uncertainty in the estimation of LF and SFRD, we explore the effect of different dust correction prescriptions by calculating the LF and SFRD using a constant dust extinction correction, A{$_{\textrm{H}\alpha}=1$} mag, a luminosity-dependent correction, and a stellar-mass dependent correction. The resulting \halpha LFs are well fitted using Schechter functions with best-fit parameters: L$^*=10^{42.24}$ erg s$^{-1}$, $\phi^*=10^{-2.85}$ Mpc$^{-3}$, $\alpha = -1.62$ for constant dust correction, L$^*=10^{42.31}$ erg s$^{-1}$, $\phi^*=10^{-2.8}$ Mpc$^{-3}$, $\alpha=-1.39$ for luminosity-dependent dust correction, and L$^*=10^{42.36}$ erg s$^{-1}$, $\phi^*=10^{-2.91}$ Mpc$^{-3}$, $\alpha = -1.48$, for stellar mass-dependent dust correction. The deep and wide nature of the DAWN survey effectively samples \halpha emitters over a wide range of luminosities, thereby providing better constraints on both the faint and bright end of the LF. Also, the SFRD estimates $\rho_{\textrm{SFR}}=10^{-1.39}$ M$_{\odot}$yr$^{-1}$Mpc$^{-3}$ (constant dust correction), $\rho_{\textrm{SFR}}=10^{-1.47}$ M$_{\odot}$yr$^{-1}$Mpc$^{-3}$ (luminosity-dependent dust correction), and $\rho_{\textrm{SFR}}=10^{-1.49}$ M$_{\odot}$yr$^{-1}$Mpc$^{-3}$ (stellar mass-dependent dust correction) are in good agreement with the evolution of SFRD across redshifts ($0 < z < 2$) seen from previous \halpha surveys.
\end{abstract}


\defcitealias{Coughlin2018}{C18}

\section{Introduction} \label{sec:intro}
Mapping the rate at which gas is transformed into stars in galaxies is a key component in understanding galaxy evolution. The spectrum of a galaxy contains emission features which indicate the underlying stellar populations' mass, age, and metallicity \citep{Madau2014}. Young and massive stars contribute most to the light emitted from a galaxy whereas older and fainter stellar populations make up most of the total stellar mass in a galaxy. 

Measuring the rate of star formation (SFR) in galaxies at different epochs is essential in understanding the star formation history of our universe. Several observational tracers of SFR exist including rest-frame UV, IR, radio and prominent nebular emission lines such as H$\alpha$ \citep{Kennicutt2012,Madau2014}. Young and short-lived massive stars produce copious amounts of UV emission which is absorbed by surrounding gas and dust in the galaxy; the ionized gas re-emits this in the form of nebular emission lines (such as Ly$\alpha$, H$\alpha$, [O\textsc{iii}]$_{\lambda\lambda4959,5007}$) whereas the heated dust produces continuum emission in the infrared. 

Among the tracers, the \halpha emission line is considered as one of the best indicators of SFR given that (1) the emission arises primarily due to photoionization of H\textsc{ii} regions by young, massive stars, (2) \halpha is less affected by dust extinction than UV continuum or bluer lines, and (3) \halpha is readily observed in the optical and near-IR up to $z\sim2$. Also, the relation between \halpha luminosity and SFR is relatively well calibrated \citep{Kennicutt1998,Kennicutt2012}. Like any other SFR indicator, the \halpha line is also affected by systematics, mainly due to dust attenuation, which can significantly impact the accuracy of SFR density (SFRD) estimates; however, dust extinction can be corrected to a reasonable extent. H$\alpha$ is also a valuable redshift tracer, and therefore of great interest for future space-based missions such as Euclid \citep{Laureijs2011}, WFIRST \citep{Spergel2015} and other ground-based surveys.

Early \halpha surveys have measured the luminosity function (LF) and SFRD in the local universe, $z < 0.5$ \citep[e.g.,][]{Gallego1995,Brinchmann2004,Nakamura2004,Hanish2006}, but most of these surveys used a relatively smaller sample of emission-line galaxies. However, with the advent of better instrumentation in the optical and near-IR regime, many surveys have been able to detect larger sample of \halpha emitters (at least by an order of magnitude) and have extended \halpha studies to earlier cosmic times \citep[e.g.,][]{Geach2008,Hayes2010,Sobral2013}. Most of these studies are mainly based on spectroscopic observations using continuum-selected galaxies from large surveys such as SDSS \citep[e.g.,][]{Brinchmann2004,Nakamura2004}, or grism spectroscopy \citep[e.g.,][]{Pirzkal2004,Xu2007,Straughn2009,Brammer2012,Pirzkal2013,Colbert2013,Malhotra2015,Pirzkal2018}, or narrow-band imaging \citep[e.g.,][]{Ly2007,Shioya2008,Villar2008,Dale2010,Sobral2013}.

Narrow-band (NB) imaging surveys have been able to study large samples of emission-line galaxies, thanks to the wide-format optical and near-IR cameras. This technique has several advantages: (1) narrow-band filters are able to detect emission-line galaxies preferentially, (2) they exhibit weak dependence on continuum luminosity, and (3) they probe sources of multiple emission-line types, each across a fairly narrow range of redshifts. NB surveys for \halpha have been carried out at various redshifts between $0< z <2.5$ where the \halpha line shifts from optical to near-IR regime with increasing redshift. However, it is particularly challenging to conduct surveys in the near-IR domain since the night sky at these wavelengths is dominated by narrow OH emission lines. NB surveys in the recent past which have probed large samples of \halpha emitters at redshifts $z > 0.4$ include HiZELS (\citealt{Sobral2013}; $z\sim0.4, 0.84, 1.47, 2.23$), NewH$\alpha$ (\citealt{Ly2011}; $z\sim0.8$) and \cite{Villar2008} at $z\sim0.84$.

The Deep And Wide Narrow-band (DAWN) survey is a near-infrared imaging survey that was carried out using the 4m Mayall telescope at Kitt Peak National Observatory (KPNO) in Arizona, USA. Three deep fields (COSMOS, UDS, EGS) were observed with a total exposure time of over 65 hours each and two other fields (CFHTLS-D4 and MACS0717) were observed for a total exposure time of over 20 hours each, using a narrow-band filter at 1.06$\mu$m on the NOAO Extremely Wide-Field InfraRed Imager (NEWFIRM; \citealt{Probst2004,Probst2008}). In addition, shallow exposures ($\sim$1-3 hours) of eight flanking regions around the deep COSMOS region were also obtained with an aim to detect larger number of bright emission-line sources across the field.

Using DAWN, various types of emission-line galaxies at different epochs can be selected and studied. In this paper, we have used DAWN primarily to study H$\alpha$ emitters at $z\sim0.62$. Complementing previous NB surveys of \halpha at nearby redshifts, this survey fills the void between $0.5 < z < 0.8$ by adding new measurements of the H$\alpha$ LF and SFRD at $z\sim0.62$, thereby helping us better understand the evolution of star formation across cosmic timescales. The previous DAWN \halpha result \citep[][hereafter C18]{Coughlin2018} laid emphasis on extending the LF to fainter luminosities and providing tighter constraints compared to other LFs at $z > 0.5$ from previous surveys. However, since the area covered was relatively small ($\sim$0.25 deg$^2$), the bright-end of the LF was not sufficiently constrained. The inclusion of flanking regions surrounding the deep COSMOS region extends the area coverage to $\sim$1.5 deg$^2$ with a co-moving volume of $\sim3.5\times10^4$ Mpc$^3$; this work improves upon \citetalias{Coughlin2018} by providing robust constraints on the bright and faint end of the H$\alpha$ LF as well as the SFRD estimate at $z\sim0.62$.

The paper is organized as follows. Section \ref{sec:obs} describes the DAWN observations and data reduction process including photometric calibration and source extraction. In section \ref{sec:sample}, we discuss the selection criteria for our emission-line galaxy sample, and the selection of \halpha emitters using spectro-photometric redshift and color-color criterion. In section \ref{sec:analysis}, we calculate \halpha luminosities taking into account [N\textsc{ii}] contamination and dust attenuation; we also determine the incompleteness arising due to selection effects and compute relevant correction factors for LF calculations. Results are presented in section \ref{sec:results} including the \halpha LF and SFRD estimate at $z\sim0.62$. The main conclusions of this work are summarized in section \ref{sec:concl}. 

Throughout the paper, we have assumed $\Lambda$-CDM cosmology: $\Omega_M = 0.3$, $\Omega_{\Lambda}=0.7$ and $H_0 = 70$km s$^{-1}$ Mpc$^{-1}$, and Salpeter IMF in our calculations. All magnitudes reported in this paper are based on the AB magnitude system.

\section{Observations and Data}\label{sec:obs}

The DAWN observations were carried out using a custom-made narrow-band filter installed on the NEWFIRM instrument with the 4m Mayall telescope at the Kitt Peak National Observatory. NEWFIRM houses a mosaic of four 2K$\times$2K InSb detectors with a chip gap of 35 arcsec and an overall field-of-view of $\sim28\arcmin \times 28\arcmin$ at 0.4 arcsec/pixel. The narrow-band filter, \texttt{NB1066}, is a custom designed filter centered at 1.066$\mu$m with a FWHM of 35\AA. With a target 5$\sigma$ limiting line flux $\sim6\times10^{-18}$ erg cm$^{-2}$ s$^{-1}$, the DAWN survey was optimized for high sensitivity and large area coverage. Using the 1.06$\mu$m narrow-band filter, this survey is able to detect galaxies showing prominent emission associated with any of the strong emission lines (Ly$\alpha$, H$\alpha$, [O\textsc{iii}]$_{\lambda\lambda4959,5007}$, [O\textsc{ii}]$_{\lambda3727}$), each at different redshift. In this work, we focus on H$\alpha$ emitters at redshift $z \sim$ 0.62, which represent star-forming galaxies at a time when the universe was roughly half its current age. 

\subsection{Near-IR imaging with NEWFIRM}
\begin{figure}[!t]
	\epsscale{1.2}
	\plotone{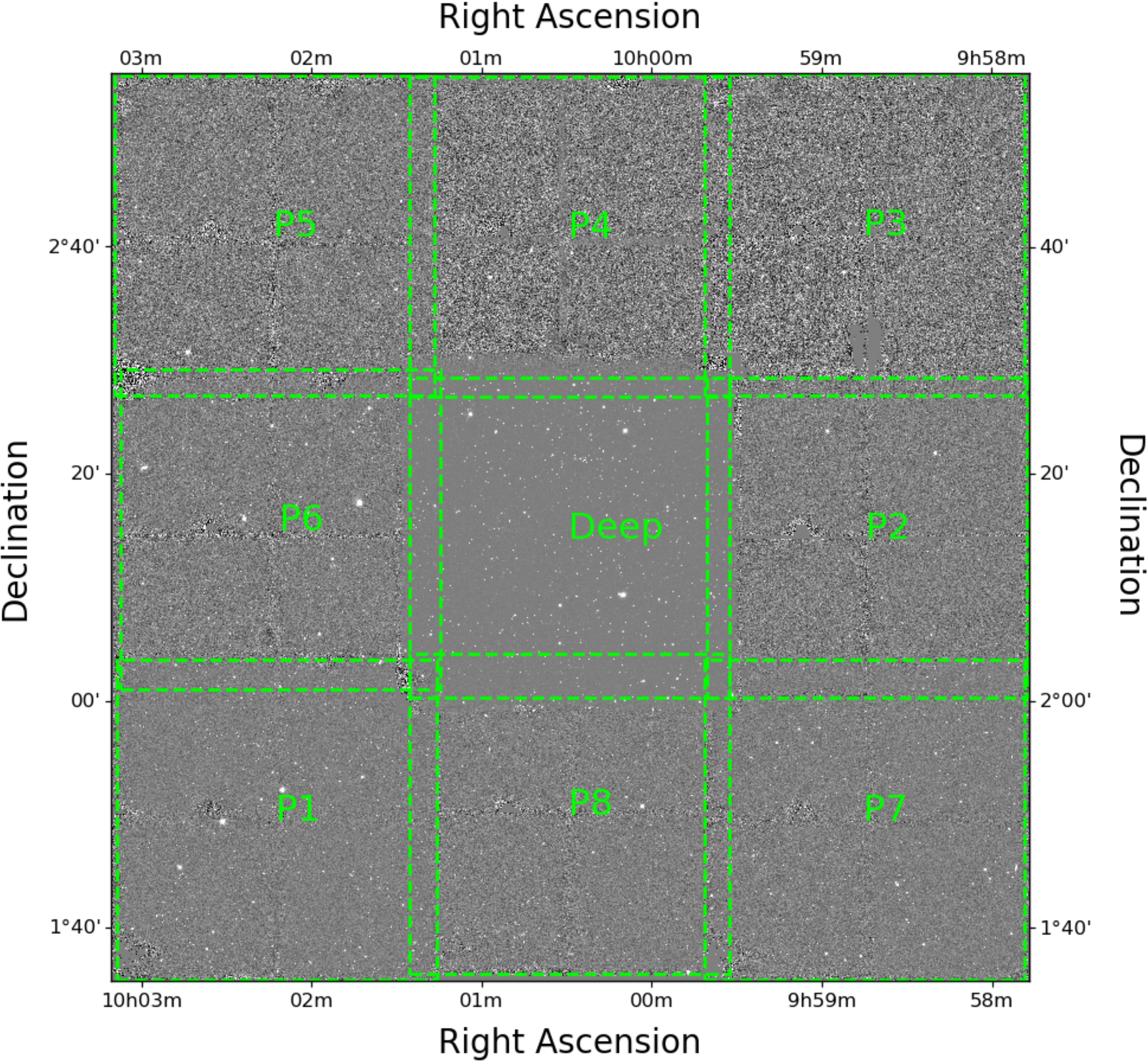}
	\caption{Coverage of the COSMOS field in DAWN survey including the \emph{deep} and \emph{flanking} (P1-P8) regions. The dimensions of each pointing is $\sim$28\arcmin x 28\arcmin. The combined \texttt{NB1066} image for the field is shown in background.}
	\label{fig:pointings}
\end{figure}

In order to better constrain the bright end of the \halpha LF, medium-deep images in eight pointings flanking the deep COSMOS region were obtained as part of the NOAO survey program 2013B-0236 (PI: Finkelstein; Stevens et al. 2019, submitted). We present an overview of these eight fields as well as the deep field in Table \ref{tab:obsCOSMOS} and Figure \ref{fig:pointings}. The dithering strategy and readout patterns followed were similar to that of \citetalias{Coughlin2018}. Full details regarding the DAWN survey will be presented in an upcoming paper (Rhoads et al., in prep).

The data reduction was performed using the NEWFIRM pipeline \citep{Swaters2009} which produced images that were calibrated, sky-subtracted, re-projected, and resampled along with their corresponding bad-pixel masks. The seeing FWHM for each of the pointings varied due to changing weather conditions across different observing nights. The final stacked images had slightly different total integration times across different flanking regions (see Table \ref{tab:obsCOSMOS}).

\subsection{Archival data and Photometry}
We used publicly available Y and J band images from the UltraVISTA survey DR3 \citep{McCracken2012} because they are substantially deeper than the NEWFIRM broad-band images that were obtained along with our narrow-band data. These images are one of the deepest near-infrared observations of the COSMOS region covering a total area of $\sim$1.5 deg$^2$, reaching 5$\sigma$ (2\arcsec\ aperture, AB) depths of $\sim$25 mag in Y and $\sim$24 mag in JHKs bands. Since VIRCAM/UltraVISTA images have a higher spatial resolution of 0.15 arcsec/pixel compared to NEWFIRM, the broad-band images were downgraded to a pixel resolution of 0.4 arcsec/pixel, using \textsc{SWARP} \citep{Bertin2002} software, to match the NEWFIRM observations. In each case, the resulting image was inspected for any evidence of astrometric mismatch by overlaying center coordinates of known bright point sources from the 2MASS catalog \citep{Skrutskie2006} and blinking between images to check for misalignment. The astrometric alignment between images matched well within a single pixel offset ($\lesssim 0.1 \arcsec$).

\begin{deluxetable}{cccCCC}[!t]
	\tablecaption{Observation summary of DAWN-COSMOS fields \label{tab:obsCOSMOS}}
	\tablecolumns{6}
	\tablenum{1}
	\tablewidth{0pt}
	\tablehead{
		\colhead{Pointing} & \colhead{RA} & \colhead{Dec.} & \colhead{Int. Time} & \colhead{FWHM} & \colhead{Depth\tablenotemark{a}} \\
		\colhead{} & \colhead{(J2000)} & \colhead{(J2000)} & \colhead{(hr)} & \colhead{(arcsec)} & \colhead{($5\sigma$, AB)}
	}
	\startdata
	Deep & 10:00:30 & +02:14:45 & 81 & 1.4 & 23.6 \\
	P1 & 10:02:13 & +01:49:27 & 3 & 1.4 & 22.1 \\
	P2 & 09:58:45 & +02:14:37 & 1 & 1.5 & 20.1 \\
	P3 & 09:58:46 & +02:41:17 & 1.5	& 1.4 & 20.5 \\
	P4 & 10:00:30 & +02:41:15 & 1.67 & 1.2 & 20.3 \\
	P5 & 10:02:14 & +02:41:02 & 2 & 1.3 & 21.0 \\
	P6 & 10:02:10 & +02:15:06 & 2.5 & 1.4 & 21.6 \\
	P7 & 09:58:46 & +01:49:47 & 2 & 1.3 & 21.9 \\
	P8 & 10:00:30 & +01:50:25 & 2 & 1.2 & 22.0 \\
	\enddata
	\tablenotetext{a}{ Depth measurements were based on 2\arcsec\ apertures.}
\end{deluxetable}

\subsection{DAWN Survey}
\subsubsection{Photometric calibration}
In order to facilitate an accurate comparison between the narrow-band and broad-band images, all images were calibrated using an artificial 1.066 $\mu$m continuum magnitude based on the interpolation between the Y- and J-band magnitudes as presented in \citetalias{Coughlin2018}. For this purpose, the UltraVISTA K-selected Catalog v4.1 \citep{Muzzin2013} was used which provides photometry for sources in YJHKs broad-bands. The image calibration was performed using only sources with NB magnitudes fainter than 15, which are bright but not saturated, and brighter than 19, which includes those detected with high signal-to-noise (SNR$>20$). Thereafter, the zero-point of all images were set to a magnitude of 30 (AB). This ensured that the median NB-excess (Y -- \texttt{NB1066}) color for unsaturated and bright sources (typically between 17-21 mag) is around zero. In order to maintain uniformity throughout, flux measurements were made using a 2\arcsec\ diameter aperture across all images. An aperture correction of -0.35 mag was applied to account for the differences in seeing.

\subsubsection{Source detection and multi-wavlength photometry}\label{sec:sextractor}
For all images, detection and extraction of sources was performed using SourceExtractor (also known as \textsc{SExtractor}) \citep{Bertin1996}. Photometry was measured using 2\arcsec\ diameter apertures with \textsc{SExtractor} run in dual-mode where narrow-band image (\texttt{NB1066}) was used as the detection image in each case while photometry was measured for Y and \texttt{NB1066}. The \textsc{SExtractor} parameters configuration used for source detection and extraction were similar to those used by \citetalias{Coughlin2018}. 

We measured the $5\sigma$ depth (as mentioned in Table \ref{tab:obsCOSMOS}) using random empty aperture (2\arcsec\ diameter) measurements of the background for each \texttt{NB1066} image. Care was taken to avoid positions where sources with SNR $\geqslant3\sigma$ are detected as well as the masked regions. A depth measurement of this kind takes into account the correlated background noise which provides a robust estimate of the noise compared to those given by \textsc{SExtractor}. However, this is also a conservative upper-limit of the noise since, occasionally, some measurements might include faint sources below the survey detection thresholds. 

\begin{figure*}
	\centering
	\epsscale{1.}
	\plotone{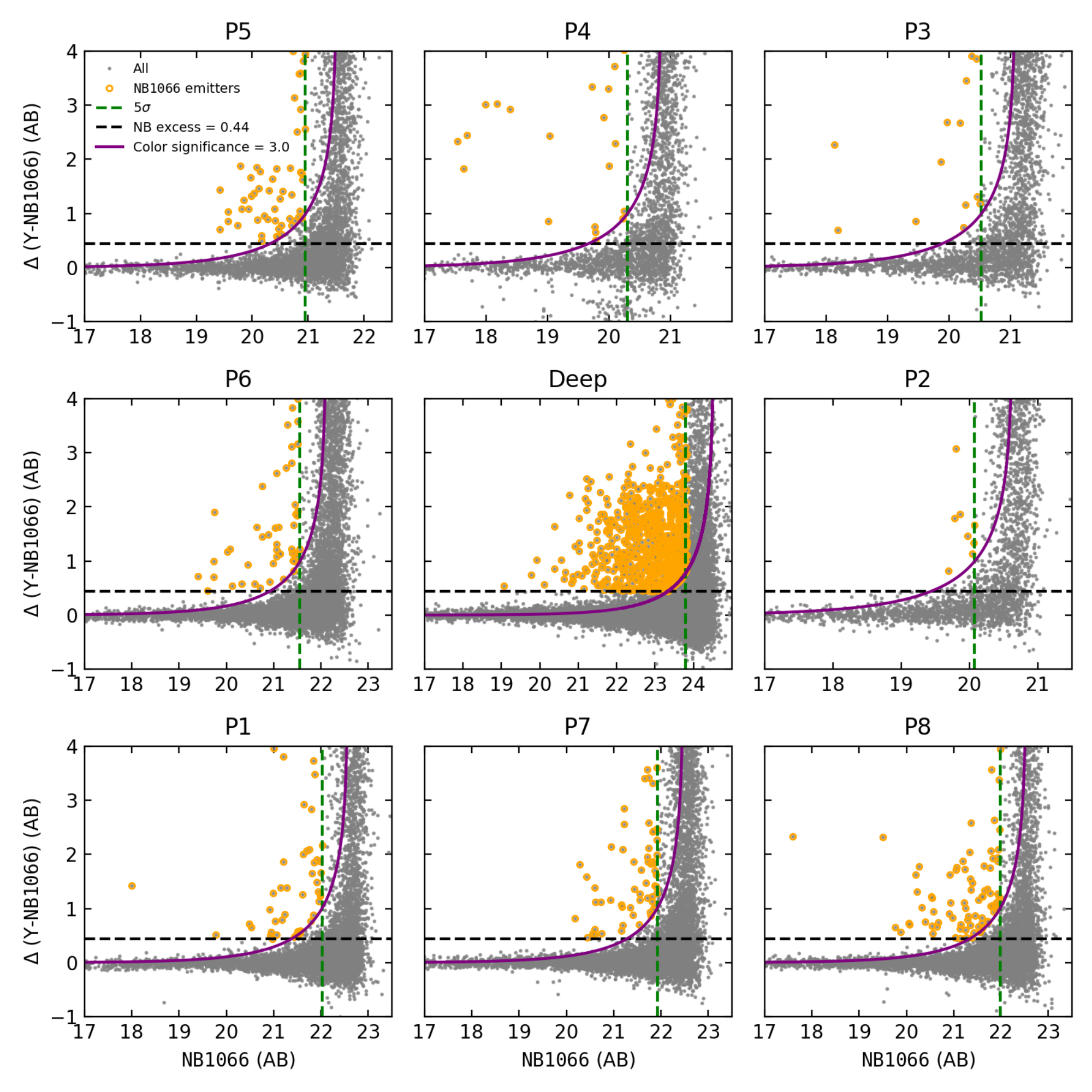}
	\caption{NB excess (Y-\texttt{NB1066}) as a function of \texttt{NB1066} magnitude used to select emission-line candidates in the deep and flanking regions. In each case, all NB detections (\emph{gray}) as well as the selected emission-line candidates (\emph{orange}) are shown. The selection criteria employed (as detailed in Section \ref{sec:sample}) includes a detection limit of SNR $\approx 5$ (\emph{vertical green dashed-line}), NB excess $\geqslant 0.44$ mag (\emph{horizontal black dashed-line}), and a minimum color significance of 3 (\emph{purple line}). }
	\label{fig:excess}
\end{figure*}

\section{Sample selection}\label{sec:sample}
In order to select emission-line objects from our source catalog, three main criteria were employed. Firstly, in each \texttt{NB1066} image, only sources with SNR $\geqslant 5$ were considered for further analysis. Considering \textsc{SExtractor} errors to be a lower estimate of the noise, given that they do not account for the correlated background noise, we scaled up \textsc{SExtractor} errors by 20\% based on the noise derived from the random empty aperture measurements in Section \ref{sec:sextractor}. This ensures that the selected candidate emission-line sources are robust detections. 

Potential emitters were selected based on their (Y -- \texttt{NB1066}) color and their significance relative to the general scatter of non-emitters with positive colors, similar to the methods employed in previous studies \citep{Villar2008, Ly2011, Sobral2013}. For any source to be considered as a line-emitter, it should be considerably brighter in the narrow-band image compared to the broad-band image. Quantitatively, our requirement was that the flux ratio of \texttt{NB1066} and Y-band detections should be, 
\begin{equation}
    \frac{f(\texttt{NB1066})}{f(\text{Y})} \geqslant 1.5
\end{equation}
This corresponds to an observer-frame equivalent width (EW$_{\text{obs}}$) of 18\AA\ at $z \sim 0.62$.

In addition, the (Y -- \texttt{NB1066}) color excess should be significant so that the sample is not dominated by errors in the photometry. Adhering to typical thresholds used in previous surveys \citep[e.g.,][]{Ly2011,Sobral2013}, the color excess significance for true emitters should be,
\begin{equation}
    \frac{f(\texttt{NB1066})-f(\text{Y})}{\sqrt{\sigma_{\texttt{NB1066}}^2+\sigma_\text{Y}^2}} \geqslant 3
\end{equation}
where $f(\texttt{NB1066})$, $f(\text{Y})$ are the flux densities and $\sigma_{\texttt{NB1066}}^2$, $\sigma_\text{Y}^2$ are the flux errors in \texttt{NB1066} and Y-band, respectively.

On applying all of the above criteria, we found 389 emission-line sources across all flanking regions put together and 774 sources in the deep region (Figure \ref{fig:excess}). These candidate line-emitters were visually inspected in \texttt{NB1066} as well as Y-band to remove artefacts/spurious objects or sources with artificially boosted fluxes due to the presence of halos of bright stars or neighboring noisy regions. With efficient masking of bad regions including instrument chip-gaps, only $\sim$2\% of the sources had to be excluded. The final sample of candidate line-emitters includes 1163 sources. 

\begin{figure}
	\epsscale{1.2}
	\plotone{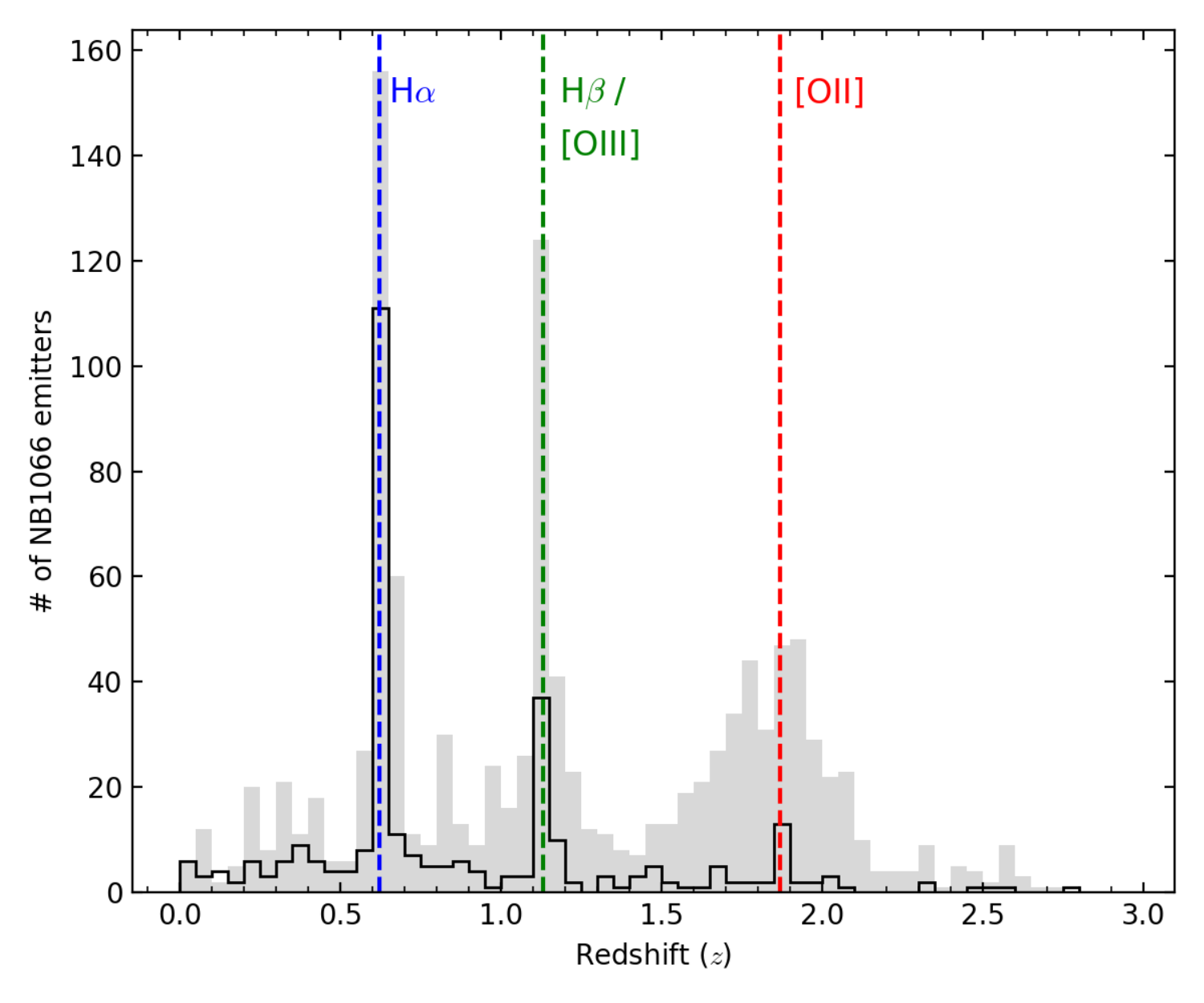}
	\caption{Redshift distribution for \texttt{NB1066}-excess selected sources using photometric redshifts (\emph{gray-shaded histogram}) from \cite{Laigle2016} and spectroscopic redshifts (\emph{black histogram}) from a compilation of various surveys (mentioned in Section \ref{sec:select}). Peaks in the distribution correspond to the redshifts at which prominent emission-line sources are detected in our survey and are labelled accordingly (\emph{dashed lines}).}
	\label{fig:redshift_dist}
\end{figure}

\begin{figure*}
	\epsscale{1.15}
	\plotone{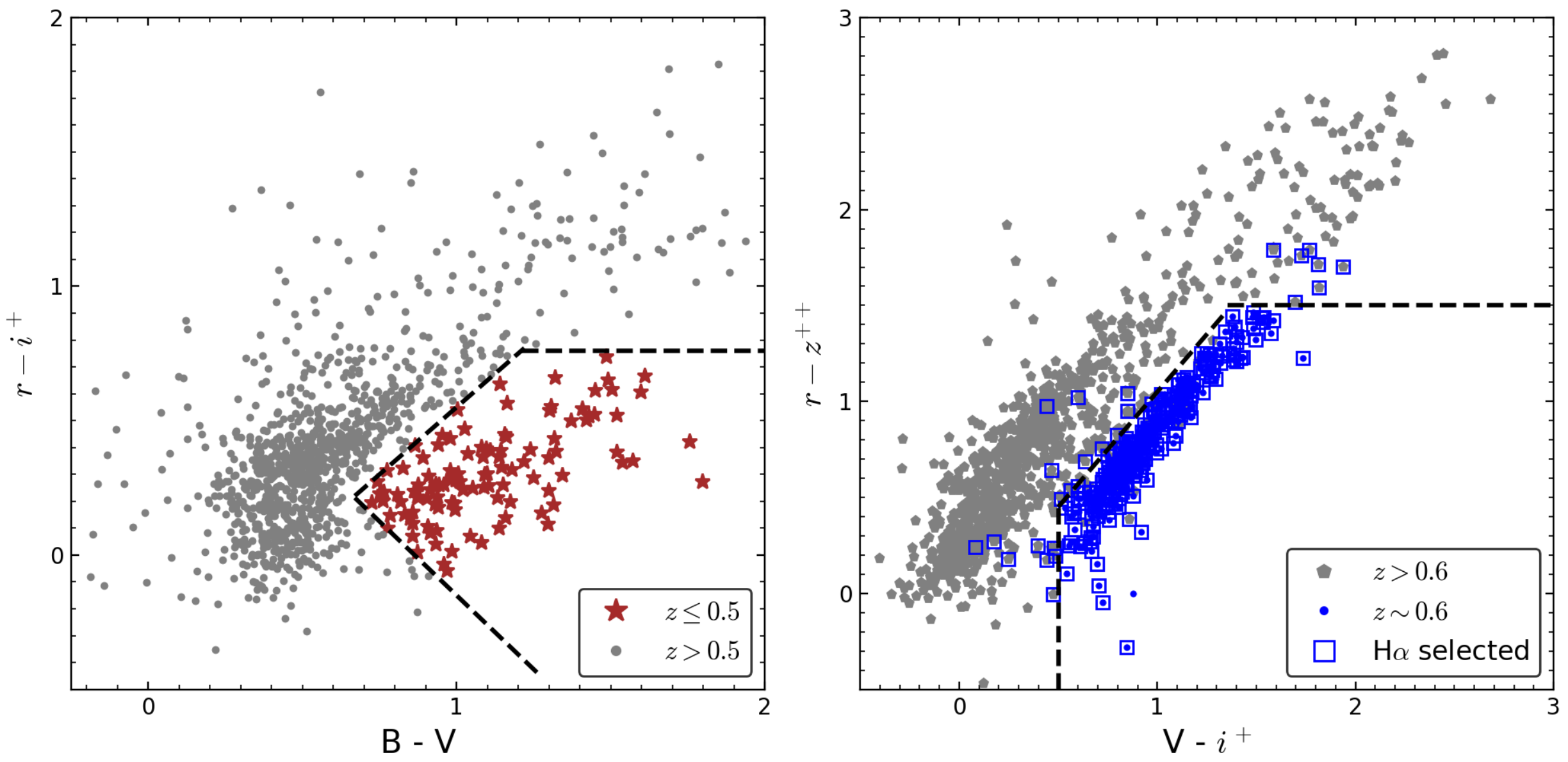}
	\caption{Color-color criteria used in the selection of H$\alpha$ emitters from the sample of candidate line-emitters. On the left, (B - V) vs. ($r-i^+$) colors help separate low-redshift sources ($z < 0.5$) from rest of the sample whereas (V - $i^+$) vs. ($r-z^{++}$)  colors, on the right, provide clear separation between $z \sim 0.62$ and the high-redshift sources. H$\alpha$ selected sources using all the criteria given in Sec. \ref{sec:select} are also indicated.}
	\label{fig:color_color}
\end{figure*}

\subsection{Selection of H$\alpha$ emitters} \label{sec:select}
The sample of candidate line-emitters includes various kinds of line-emitters such as H$\alpha$, H$\beta$/[O{\sc iii}]$_{\lambda\lambda4959,5007}$ and [O{\sc ii}]$_{\lambda3727}$. The nature of each source, in terms of their line emission, can be determined using several methods. A robust confirmation would be a match with available spectroscopic-redshift catalogs. However, because of a lack of large number of spectroscopic confirmations, a match with photometric-redshift catalogs would be the next best means to categorize these emission-line sources. Narrow-band filters are designed in such a way that they are expected to detect line-emitters exquisitely, which have strong and narrow emission-lines, potentially with little to no continuum detected in the narrow-band. For sources with faint continuum, it is possible that photometric-redshifts might be unreliable or even non-existent. Therefore, for such sources, a color-color calibration based on spectroscopically confirmed sources (and their broad-band photometry) can be used for the classification. 

For spectroscopic matches, we use a master catalog of spectroscopic redshifts compiled from various past surveys covering the COSMOS region: zCOSMOS \citep{Lilly2009}, 10K-DEIMOS \citep{Hasinger2018}, 3D-HST (\citet{Brammer2012,Momcheva2016}), VLT/FORS2  observations \citep{Comparat2015}, C3R2 \citep{Masters2017}, FMOS-COSMOS \citep{Silverman2015}, GEEC2 \citep{Balogh2014}, COSMOS-[O\textsc{ii}] \citep{Kaasinen2017}, LEGA-C \citep{Straatman2018}, MOSDEF \citep{Kriek2015}, PRIMUS \citep{Coil2011,Cool2013}, MMT/Hectospec observations \citep{Prescott2006}, and Magellan/IMACS observations\citep{Trump2009}. All sources with redshifts in the range $0.6 \leqslant z_{spec} \leqslant 0.65$ were selected as H$\alpha$ emitters irrespective of their quality flag since NB-excess selected sources are a reaffirmation to the measured spectroscopic redshifts. 

Using multi-wavelength observations from UV to near-IR, the COSMOS2015 \citep{Laigle2016} catalog contains one of the largest compilation of photometric redshifts in the $\sim$2 deg$^2$ COSMOS field. Given the uncertainties associated with photometric-redshifts, all sources with redshifts in the range $0.57 \leqslant z_{phot} \leqslant 0.67$ were selected as H$\alpha$ emitters. Figure \ref{fig:redshift_dist} shows the spectroscopic and photometric-redshift distribution for all NB-excess selected sources. In both the distributions, there are well-defined peaks at $z \sim $ 0.62, 1.13 and 1.86, corresponding to the line-emitters H$\alpha$, H$\beta$/[O{\sc iii}] and [O{\sc ii}] detected by our \texttt{NB1066} filter, respectively. Since the COSMOS2015 catalog is based on a stacked zYJHKs image, the catalog is highly complete relative to our H-alpha sample, given that our \texttt{NB1066} filter overlaps with Y-band. Out of the 1163 candidate line-emitters, $\sim$ 98\% of the sample contained redshift estimates, either photometric or spectroscopic, and in some cases, both.

Apart from the redshift-based characterization, we also used broadband photometry available from \cite{Laigle2016} to categorize sources based on color-color selection criteria. For our sample, we used the following color-color criteria to select H$\alpha$ emitters (Figure \ref{fig:color_color}):
\begin{equation}
	\begin{gathered}
		(r-i^+) < 0.75\ \text{and}\ (r-i^+) < (\text{B--V})-0.45 \\ \text{and}\ (r-i^+) > -1.1(\text{B--V})+0.95
	\end{gathered}
\end{equation}

\begin{equation}
\begin{gathered}
(\text{V}-i^+) > 0.5\ \text{and}\ (r-z^{++}) < 1.5 \\ \text{and}\ (r-z^{++}) < 1.2(\text{V}-i^+)-0.15
\end{gathered}
\end{equation}

The BV$ri^+$ criteria separates low-redshift sources ($z < 0.5$) from all other higher redshift sources in the sample. After excluding these low-redshift sources, the V$i^+rz^{++}$ criteria is used to separate $z\sim0.62$ sources, which are mostly H$\alpha$, from other high-redshift sources (mostly, [O{\sc iii}] and [O{\sc ii}]) in the sample. A drawback of this method is that some interlopers might get wrongly selected as H$\alpha$ emitters and some genuine H$\alpha$ sources might lie outside our color-color selection region. However, we can measure the fraction of contaminating as well as missed sources using spectroscopically confirmed sources, and the total contamination fraction remains relatively low ($<10\%$). Finally, 241 sources were selected as H$\alpha$ emitters with 111 sources selected based on spectroscopic redshifts, 110 sources using photometric redshifts and 20 unique sources using the broad-band color-color criteria.

\section{Analysis} \label{sec:analysis}

\subsection{H$\alpha$ luminosities}
Using \texttt{NB1066} and Y-band flux densities, the emission-line fluxes ($F_\text{L}$) and observed equivalent width (EW$_{\text{obs}}$) for our H$\alpha$ sample were calculated as follows. 
\begin{equation}
   F_\text{L} = \Delta\text{NB} \left(\frac{f_\text{NB}-f_\text{Y}}{1-(\Delta\text{NB}/\Delta\text{Y})}\right),
\end{equation}
\begin{equation}
    EW_{\text{obs}} = \Delta\text{NB} \left(\frac{f_\text{NB}-f_\text{Y}}{f_\text{Y}-f_\text{NB}(\Delta\text{NB}/\Delta\text{Y})}\right)
\end{equation}
where $\Delta${NB}, $\Delta$Y are the filter widths (FWHM in \AA), and $f_\text{NB}$, $f_\text{Y}$ are the flux densities (erg s$^{-1}$ cm$^{-2}$ \AA$^{-1}$) for \texttt{NB1066} and Y-band, respectively. The corresponding line luminosities are derived assuming $z = 0.62$, which is the median redshift of our H$\alpha$ sample.

The intrinsic H$\alpha$ luminosity can be derived using the observed line luminosity after correcting for contamination due to adjacent [N{\sc ii}]$_{\lambda\lambda6548,6584}$ lines, as well as attenuation due to dust.

\subsection{{\sc [Nii]} contamination}
For typical L$_*$ galaxies in the nearby universe, past surveys have adopted corrections based on the typical H$\alpha$/[N{\sc ii}] flux ratio of 2.3 \citep{Kennicutt1992, Gallego1997}. However, recent narrow-band surveys \citep{Villar2008, Ly2011, Sobral2013} have adopted EW-dependent corrections based on the mean relationship between rest-frame EW of H$\alpha$+[N{\sc ii}]$\lambda$6583 and the H$\alpha$/[N{\sc ii}] ratio. 

Unlike previous surveys, the narrow-band filter, \texttt{NB1066}, in DAWN is relatively narrow such that, for any \halpha source at $z \sim 0.62$, only one of the [N{\sc ii}] lines is expected to be contaminating the H$\alpha$ line flux. For example, when \halpha is detected on the bluer side of the filter, the redder [N\textsc{ii}]$_{\lambda6584}$ line is expected at the wings of the filter ($< 50\%$ transmission). However, if \halpha is detected on the redder part, the bluer [N\textsc{ii}]$_{\lambda6548}$ line should fall in a reasonably transmissive portion of the filter. Owing to a lack of spectroscopic redshift for each source in our H$\alpha$ sample, deriving individual correction is fairly difficult given that it is highly impractical to determine which one of the [N{\sc ii}] lines is responsible for contamination in each source. Therefore, the necessary corrections in this case are intrinsically different compared to other surveys. 

For these aforementioned reasons, we derive a luminosity-dependent [N\textsc{ii}] correction as follows:  We generate a mock galaxy sample with luminosities and redshifts based on the observed luminosity distribution and a uniform redshift distribution comprising redshifts probed by our \texttt{NB1066} filter. Assuming a fixed [N\textsc{ii}]/\halpha value of 0.43 \citep{Kennicutt1992,Gallego1997}, we derive \halpha luminosities for all sources in the mock sample after convolving them through our \texttt{NB1066} filter curve (case `A'). In a similar way, we also derive \halpha luminosities assuming [N\textsc{ii}]/\halpha $\sim 0$ for the mock sample (case `B'). Using the \halpha luminosity distribution from case `A' and `B', we calculate their ratio as a function of luminosity, which provides an estimate of the factor by which objects have been over-counted per luminosity bin. The resulting correction factor is applied to the LFs in Sec. \ref{sec:LFs} which corrects for the presence of [N\textsc{ii}] within our \halpha sample.

\subsection{Dust attenuation} \label{sec:dust}
Dust obscuration is a significant source of uncertainty in UV and optical measurements of galaxy properties including SFR. Although H$\alpha$ emission is less affected by dust compared to the UV continuum, correcting H$\alpha$ luminosities for dust is necessary to accurately measure SFR. Ideally, dust corrections applied to galaxies should be measured individually, for example, based on Balmer decrements (H$\alpha$/H$\beta$), but that requires rest-frame optical/near-IR spectra for each galaxy \citep[e.g.,][]{Reddy2015} which is practically infeasible for large samples.

In the past, some studies \citep[e.g.,][]{Sobral2013} have adopted a simple dust correction of A$_{\text{H}\alpha}$=1 assuming that dust affects all sources in the sample equally, whereas some others \citep[e.g.,][]{Ly2011} have assumed that dust extinction in galaxies depend on their SFR/luminosity \citep{Hopkins2001} or stellar-mass \citep{GB2010} and hence apply corrections accordingly. \citet{Sobral2013} believe that the typical extinction in a galaxy need not necessarily depend on its SFR/luminosity in an absolute manner, but rather depend on the nature of the source (meaning the extent to which it is star-forming or luminous) relative to the normal star-forming galaxy at a particular epoch.

\begin{figure*}
	\epsscale{1.}
	\plotone{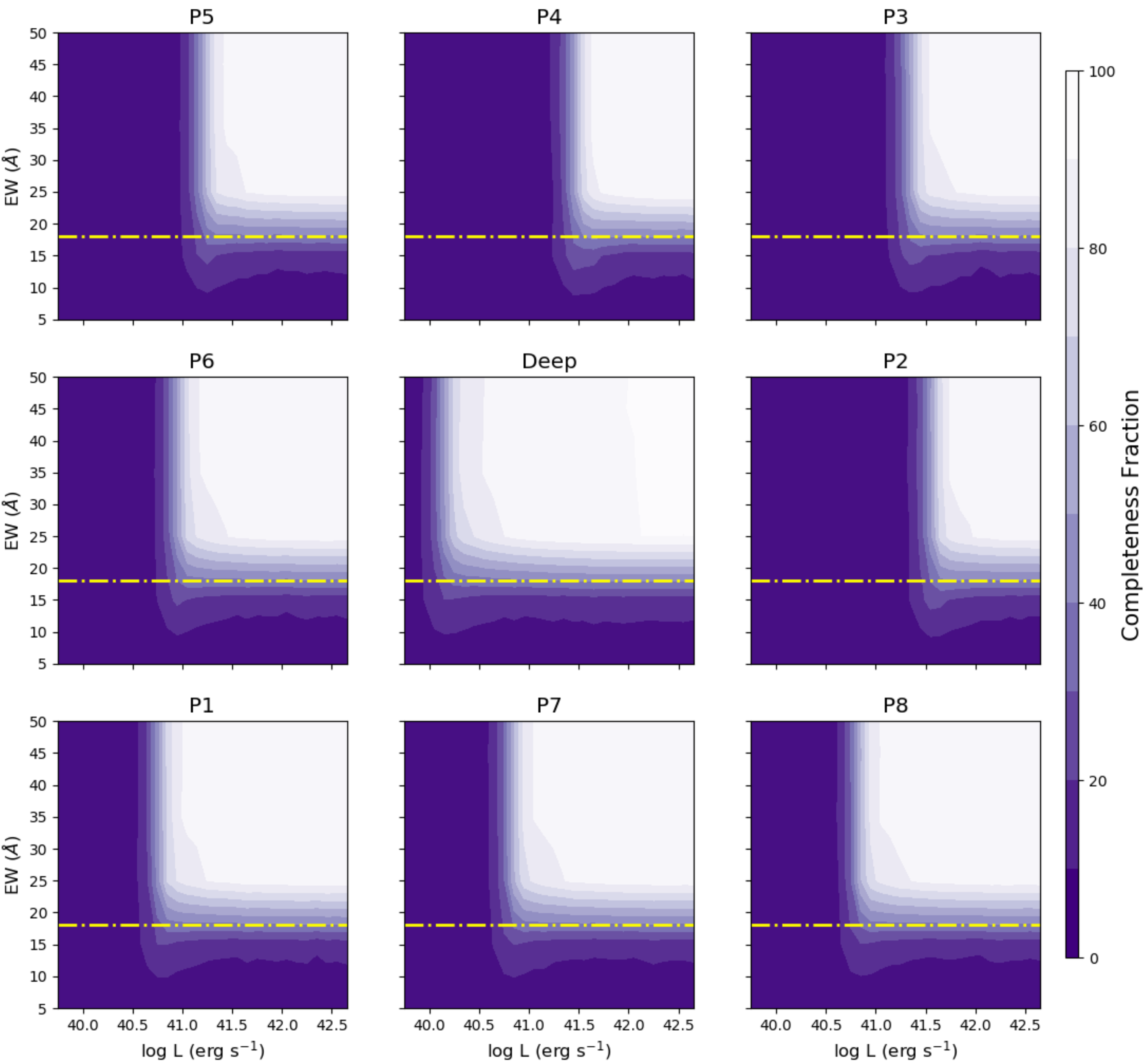}
	\caption{Completeness fraction as a function of luminosity and EW for deep and flanking regions. The dashed line represents the EW-cut adopted in this survey.}
	\label{fig:completeness_flanking}
\end{figure*}

In order to explore the effects of different dust-extinction correction on LF and SFRD, we correct our \halpha luminosities following all aforementioned prescriptions and analyze them separately hereafter. For luminosity-dependent extinction correction, we adopted the following relation given by \citet{Ly2012}:
\begin{equation}
\log(L_{\text{obs}}) = \log(L_{\text{int}}) - 2.36 \ \log\left[\frac{0.797\  \log(L_{\text{int}})-29.1}{2.86}\right],
\end{equation}
where $L_{\text{obs}}$ and $L_{\text{int}}$ are the observed and intrinsic H$\alpha$ luminosities (erg s$^{-1}$), respectively. As mentioned earlier, this extinction correction is based on the SFR-dependent formalism derived by \citet{Hopkins2001} which demonstrates that H$\alpha$ luminosity directly correlates with SFR, meaning dust reddening will be higher for sources with higher H$\alpha$ luminosity. In case of stellar mass-dependent correction, the dust extinction is computed according to the following relation given by \citet{GB2010}:
\begin{equation}
	A_{\text{H}\alpha} = 0.91 X + 0.77 X^2 +0.11 X^3 - 0.09 X^4
\end{equation}
where $X$ = log$_{10}$($M_*$/10$^{10}$ M$_\odot$).  For our \halpha sample, the stellar mass estimates available from COSMOS2015 \citep{Laigle2016} are used to derive the extinction corrections. Since several sources in our sample have stellar masses log (M$_*$/M$_\odot$) $<  8.5$, where the paramaterization does not account for such low stellar mass sources, we assume a fixed dust correction, A$_{\text{H}\alpha}$ = 0.3 mag, corresponding to the extinction correction derived for a source with log (M$_*$/M$_\odot$) $\sim 8.5$, for all such sources.

\subsection{Completeness corrections} \label{sec:comp}
Given our methods of detection and selection of H$\alpha$ emitters, we have to estimate the incompleteness arising out of this process and apply an appropriate correction for each source in our sample. Based on the procedure suggested by \citetalias{Coughlin2018}, we estimate the completeness fraction of our sample. Briefly put, artificial sources are randomly superimposed on the science image. The standard detection and selection methods are followed to determine the number of sources recovered. A comparison between the number of emission-line detected sources ($N_{\text{detected}}$) and the number of artificial ($N_{\text{artificial}}$) plus real ($N_{\text{real}}$) sources present in the image provides us with a recovery fraction for the sample. 
\begin{equation}
\kappa = \frac{N_{\text{detected}}-N_{\text{real}}}{N_{\text{artificial}}}
\end{equation}
This procedure is repeated once for each bin across a range of luminosities and EW$_{\text{obs}}$.

Owing to different image depths across the deep and flanking regions, the completeness simulation was performed individually for each region. Simulations were performed for each of the 600 bins across a luminosity range: 10$^{39.7}$ -- 10$^{42.7}$ L$\odot$ ($\Delta \sim$ 0.1 dex), and an EW$_{\text{obs}}$ range: 0 -- 200 \AA\ ($\Delta \sim$ 10 \AA). Since the total number of sources detected in the flanking regions is less than half the number detected in the deep region, the completeness simulation for flanking regions included 5,000 artificial sources whereas the simulation for deep region included 10,000 artificial sources. The completeness correction thereby computed was applied to each source, depending on its luminosity and EW$_{\text{obs}}$, within each region.

Figure \ref{fig:completeness_flanking} shows completeness fractions for the deep and flanking regions. We adopt a 20\% completeness limit for each luminosity-EW$_{\text{obs}}$ bin while applying corrections for sources in a particular region.

\begin{figure*}
	\centering
	\epsscale{1.2}
	\plotone{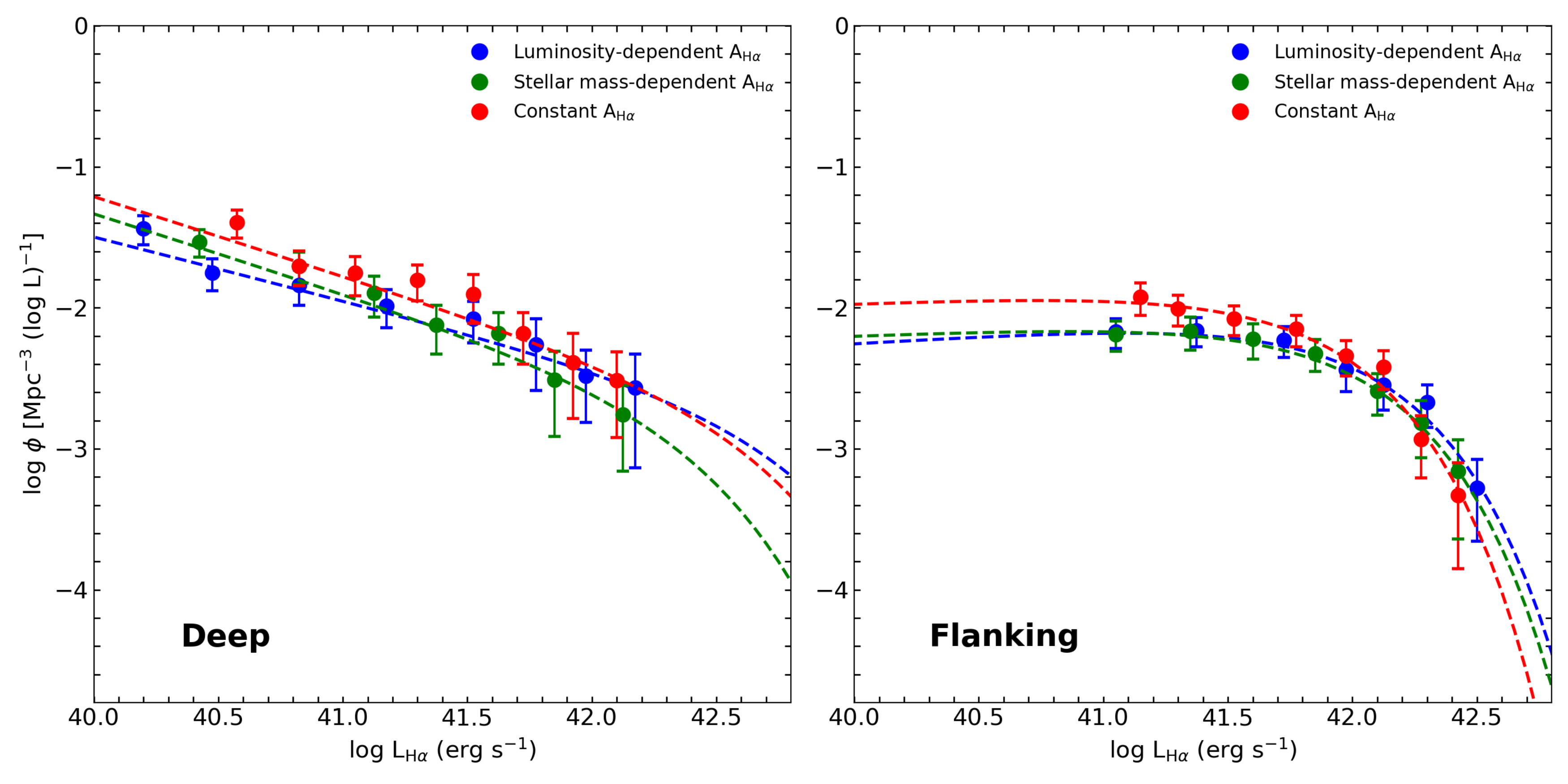}
	\caption{H$\alpha$ LF with their respective Schechter fits for the deep (\emph{left}) and flanking regions (\emph{right}). These LFs are corrected for [N\textsc{ii}] contamination, incompleteness as well dust-extinction (luminosity-dependent correction in \emph{blue}, constant correction in \emph{red}, and stellar mass-dependent correction in \emph{green})}
	\label{fig:individual-LF}
\end{figure*}

\begin{deluxetable*}{ccccCCC}[t!]
	\setlength{\tabcolsep}{10pt}
	\tablecaption{Schechter parameters of the LF and SFR density for $z \sim 0.62$ H$\alpha$ emitters in DAWN--COSMOS\label{tab:LF+SFRD}}
	\tablecolumns{7}
	\tablenum{2}
	\tablewidth{0pt}
	\tablehead{
		\colhead{Region} &\colhead{Dust correction} & \colhead{L$^*$} & \colhead{$\phi^*$} & \colhead{$\alpha$} & \colhead{log $\mathcal{L}$} & \colhead{log $\rho_{\mathrm{SFR}}$} \\
		\colhead{} & \colhead{(A$_{\text{H}\alpha}$)} & \colhead{(erg s$^{-1}$)} & \colhead{(Mpc$^{-3}$)} & \colhead{} & \colhead{(erg s$^{-1}$ Mpc$^{-3}$)} & \colhead{(M$_\odot$ yr$^{-1}$ Mpc$^{-3}$)}
	}
	\startdata
	\multirow{3}{*}{COSMOS} & Luminosity-dependent & $42.31^{+0.27}_{-0.18}$ & $-2.80^{+0.22}_{-0.34}$ & $-1.39^{+0.13}_{-0.14}$ & $39.68^{+0.06}_{-0.06}$ & $-1.47^{+0.06}_{-0.06}$\\
	& Stellar mass-dependent & $42.36^{+0.35}_{-0.13}$ & $-2.91^{+0.28}_{-0.43}$ & $-1.48^{+0.16}_{-0.14}$ & $39.69^{+0.08}_{-0.07}$ & $-1.49^{+0.08}_{-0.07}$\\
	& Constant & $42.24^{+0.39}_{-0.21}$ & $-2.85^{+0.31}_{-0.42}$ & $-1.62^{+0.18}_{-0.16}$ & $39.76^{+0.08}_{-0.09}$ & $-1.39^{+0.08}_{-0.09}$\\	
	\enddata
\end{deluxetable*}

\section{Results} \label{sec:results}

\subsection{H$\alpha$ luminosity function at $z \sim 0.62$}\label{sec:LFs}
H$\alpha$ luminosity functions for this survey are derived using the  V/V$_{max}$ method \citep{Schmidt1968}. In this work, since the sample of H$\alpha$ emitters are selected from regions of varying imaging depths, we construct and analyze LFs separately for the deep region, the shallow flanking regions, and the full DAWN COSMOS region ($\sim$ 1.5 deg$^2$). Following Sec. \ref{sec:dust}, the LFs presented hereafter are derived based on all three prescriptions of dust correction, for comparison purposes. Unless otherwise specified, the errors for each LF bin are Poissonian with an additional error of 20 percent added in quadrature to account for the uncertainty in completeness corrections.

Each LF presented in this work can be modeled based on the typical Schechter profile \citep{Schechter1976} defined as follows:
\begin{equation}
	\phi(L) dL = \phi^* \left(\frac{L}{L^*}\right)^{\alpha} \textnormal{exp}\left(-\frac{L}{L^*}\right) \left(\frac{dL}{L^*}\right)
\end{equation}
In the log form, this function can be defined as,
\begin{equation}
	\phi(L) dL = \textnormal{ln}(10)\ \phi^* \left(\frac{L}{L^*}\right)^{\alpha+1} \textnormal{exp}\left(-\frac{L}{L^*}\right) d(\textnormal{log}\ L)
\end{equation}
Many H$\alpha$ surveys in the past have been able to successfully model their LF using the Schechter function; in further sections, we show that this holds good for our H$\alpha$ LF as well. We adopt a 20\% completeness limit in terms of the H$\alpha$ LF, for all calculations hereafter. 

The best-fit Schechter parameters and their associated 1$\sigma$ uncertainties are determined using MCMC simulations based on the Metropolis-Hastings algorithm. The simulation involves the following steps: (1) An intial guess for the Schechter parameters are validated against a uniform prior ($-4 < \text{log}\ \phi^*\ \text{Mpc}^{-3} < -1$, $40 < \text{log}\ L^*\ \text{erg s}^{-1} < 44$ and $-2 < \alpha < 0$). (2) Each iteration determines the goodness of the Schechter fit to the given H$\alpha$ LF based on the $\chi^2$ statistic. (3) The entire parameter space for all Schechter parameters is explored over 500,000 iterations and their probability distributions are derived. The median and 1$\sigma$ estimates from these distributions correspond to the best-fit Schechter parameters and their errors for our H$\alpha$ LF, respectively. 

\begin{figure*}\label{fig:full-LF}
	\begin{tabular}{ll}
		{\epsscale{0.6}\plotone{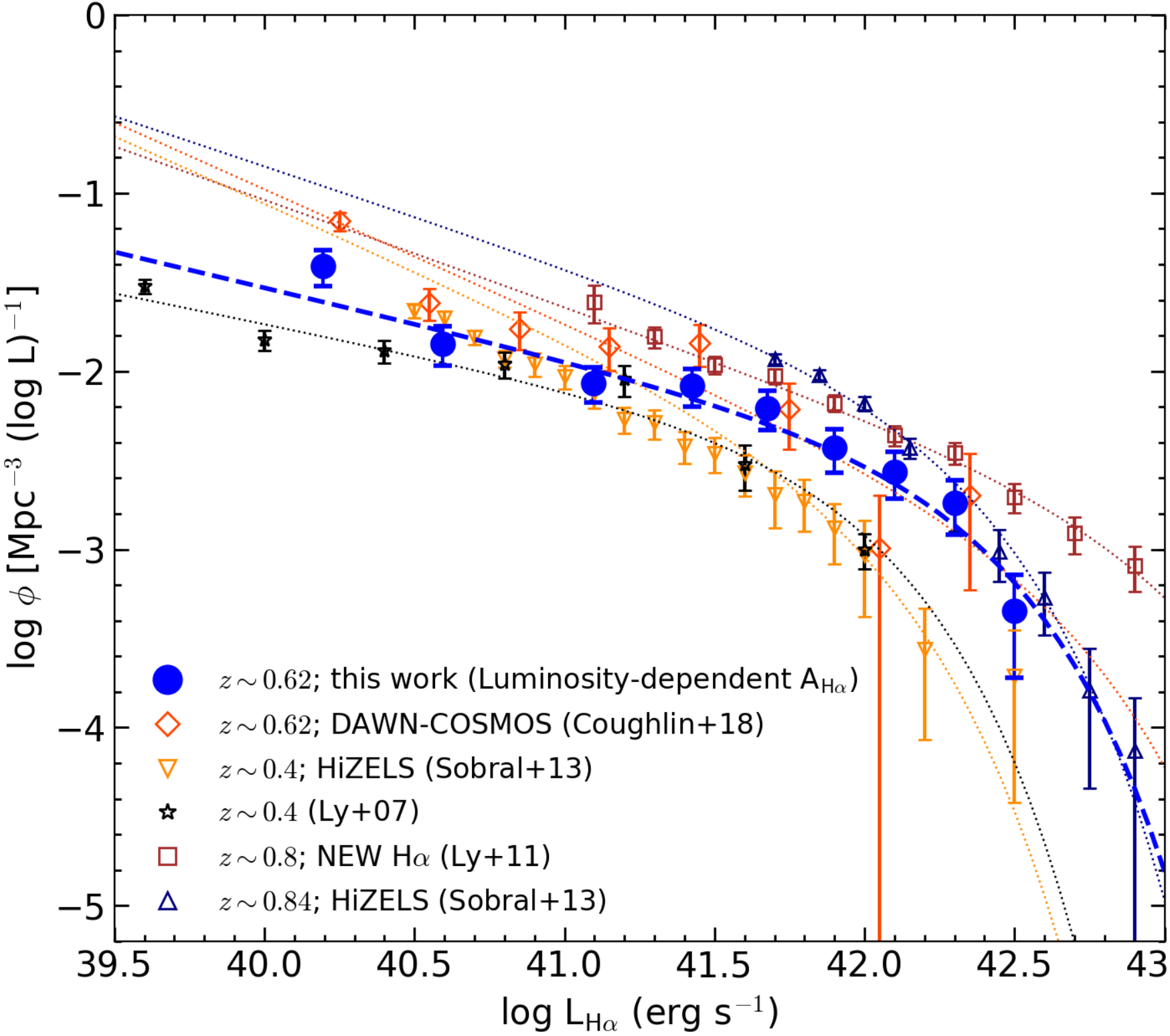}} & {\epsscale{0.575}\plotone{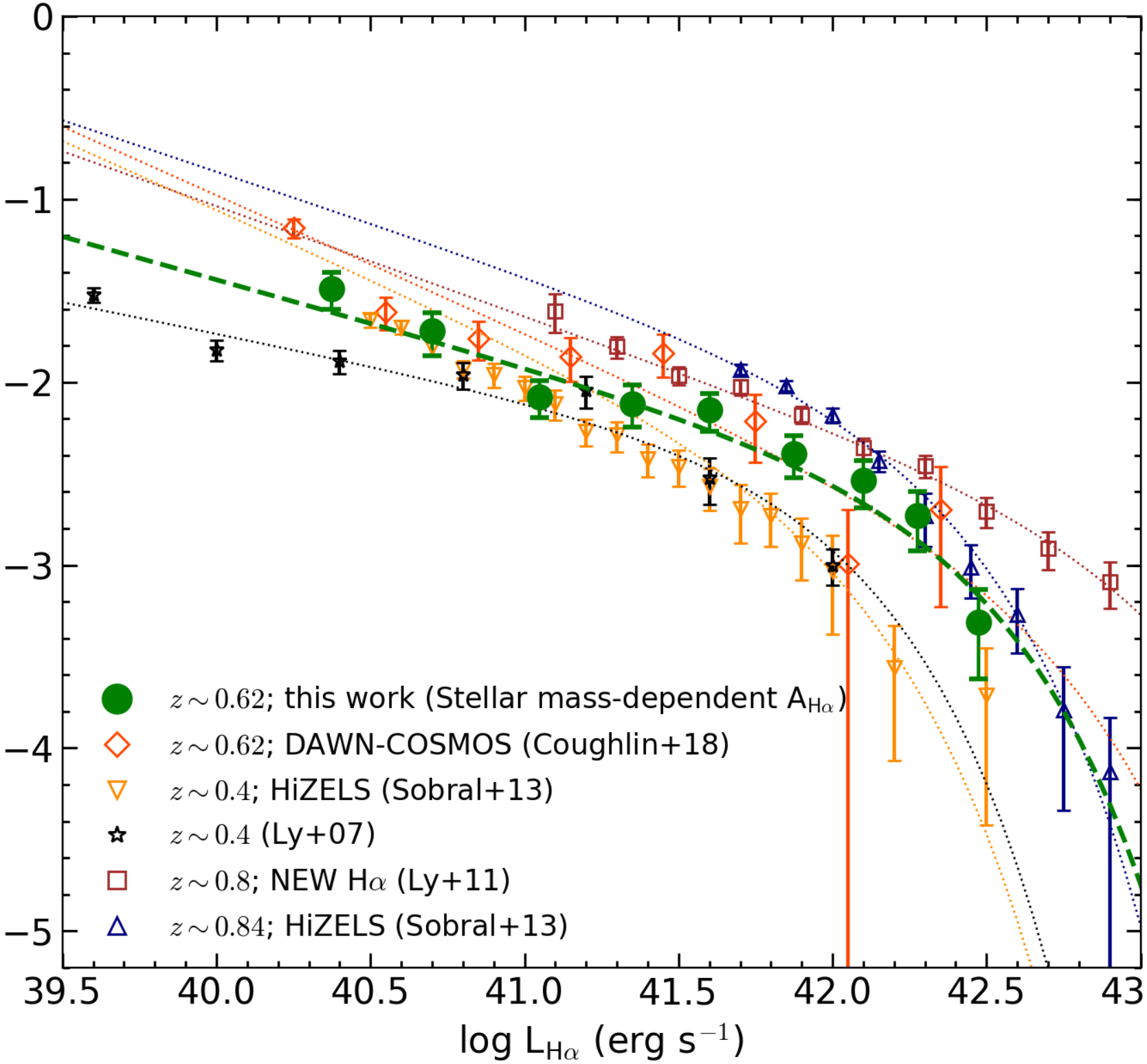}}\\ [0.5cm]
		\multicolumn{2}{c}{\epsscale{0.6}\plotone{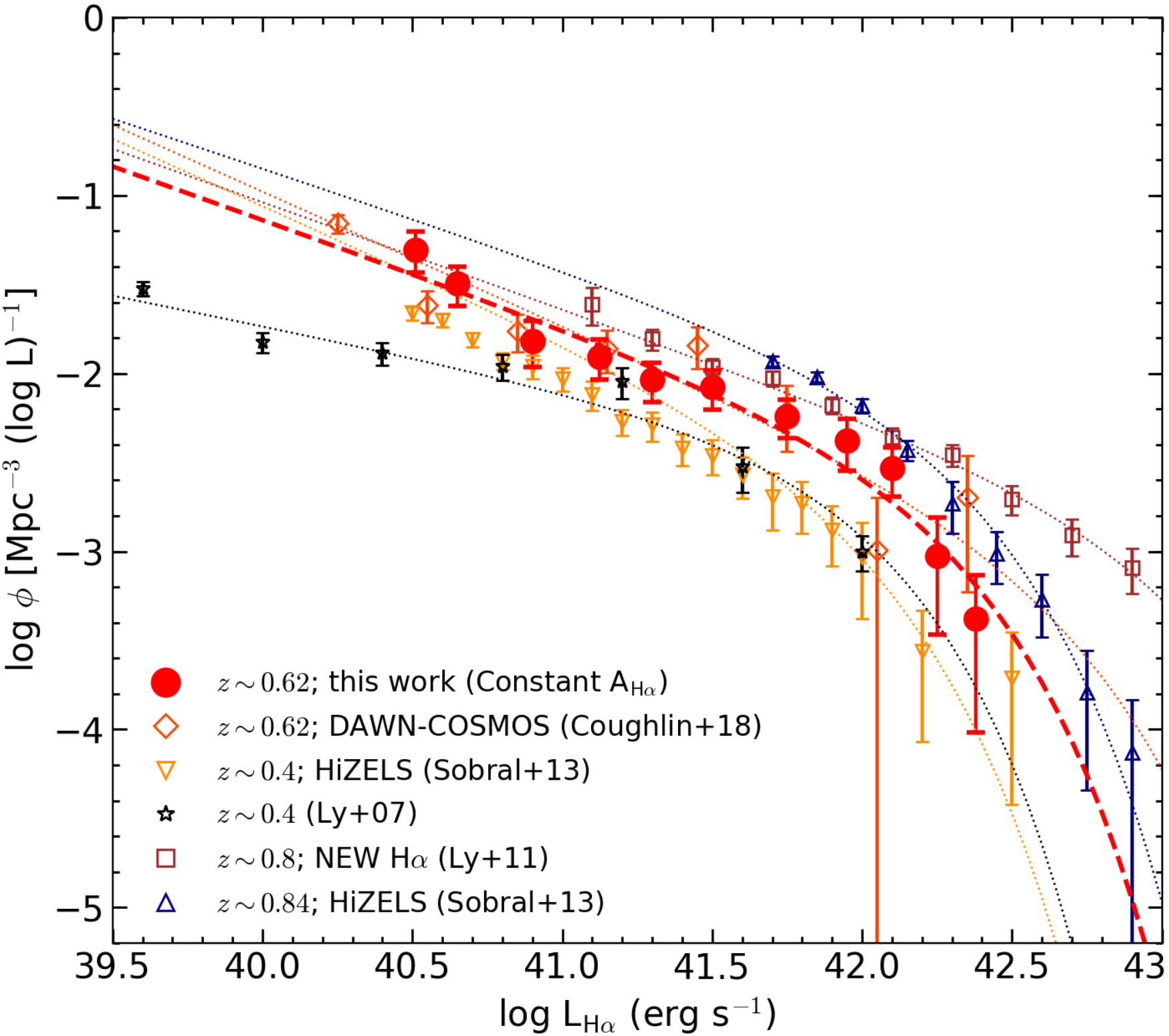}}\\
	\end{tabular}
	\caption{H$\alpha$ LFs with their corresponding Schechter fits for the full COSMOS region surveyed by DAWN and its comparison with previous \halpha surveys \citep{Ly2007,Ly2011,Sobral2013,Coughlin2018}. All LFs shown here are corrected for [N\textsc{ii}] contamination, incompleteness and dust-extinction. In case of DAWN, the LFs based on three different dust-extinction corrections are shown (\emph{top-left}: luminosity-dependent A$_{\text{H}\alpha}$, \emph{top-right}: stellar mass-dependent A$_{\text{H}\alpha}$, \emph{bottom}: constant A$_{\text{H}\alpha}$). In case of previous \halpha surveys, the LFs from \citet{Ly2007,Ly2011,Coughlin2018} are luminosity-dependent dust corrected, whereas LFs from \citet{Sobral2013} are dust-corrected assuming A$_{\text{H}\alpha}$ = 1. The DAWN LFs at $z\sim0.62$ are consistent with the LF evolution observed between redshifts, $0.4 < z < 0.84$.}
	\label{fig:full-LF}
\end{figure*}

The LFs derived using the \halpha subsample from the deep region alone are presented in Figure \ref{fig:individual-LF} (\emph{left}) which are completeness and dust-corrected. We derive three different LFs (and their respective Schechter fits) based on the three prescriptions of dust-extinction correction used in this work. In either of the LFs, it is seen that the faint-end slopes are steep and consistent with the canonical value of $\alpha$ = -1.6, which is observed among most H$\alpha$ LFs from recent NB surveys at $z \sim 0-2$ \citep[e.g.,][]{Ly2011,Sobral2013,GG2016}. However, since the volume probed by this region is relatively small ($\sim 0.25$ deg$^2$), the brighter \halpha population (L$_{\text{H}\alpha} > 10^{42}$ erg s$^{-1}$) is sparsely sampled and therefore, the bright-end of the LF is weakly constrained. 

Using the \halpha subsample from just the flanking regions, the derived LFs are as shown in Figure \ref{fig:individual-LF} (\emph{right}). In contrast to the deep region, the flanking regions put together sample significant number of bright H$\alpha$ sources (L$_{\text{H}\alpha} \geqslant 10^{42}$ erg s$^{-1}$). Unlike in the case of deep region, bright-end of the Schechter fit, which essentially depicts the break from power-law form of the Schechter function, is better constrained for these flanking-region LFs. On the contrary, the faint-end is hardly constrained owing to lack of faint luminosity sources which is expected given the shallow exposures. Therefore, the Schechter parameters for these LFs, especially the faint-end slope ($\alpha$), should not be viewed as a significant implication of this work. 

For the full \halpha sample (deep and flanking regions combined), the resulting LFs and their Schechter fits are shown in Figure \ref{fig:full-LF} in comparison with H$\alpha$ LFs from previous surveys at various redshifts, $z\sim0-2$ \citep{Ly2007,Ly2011,Sobral2013}. The best-fit Schechter parameters are given in Table \ref{tab:LF+SFRD}. Considering the empirical relation given for L$^*$ and $\phi^*$ as a function of redshift from \cite{Sobral2013}, our values are consistent with the expected value from this relation at $z\sim0.62$. We find that the characteristic luminosity as well as the normalization parameters are higher compared to those at lower redshifts. However, the faint-end slope is significantly steeper in case of the LF based on constant dust correction compared to the LF with luminosity-dependent or stellar mass-dependent dust correction. Using \citet{Hopkins2001} dust correction, previous studies have suggested that the faint-end slope tends to flatten out due to an increase in the number density on the bright-end resulting from a higher correction applied for brighter sources \cite[e.g.,][]{Villar2008,An2014}. Our faint-end slope using the same dust correction is consistent with those values. On the other hand, studies employing constant dust correction, A$_{\text{H}\alpha}=1$, observe a steeper faint-end slope, $\alpha=-1.6$, in their LFs \cite[e.g.,][]{Ly2007,Sobral2009,Sobral2013} where they argue that there is mild dependence of extinction on observed luminosity \citep{Sobral2012} and a median correction of $\sim$1 mag holds good on an average for large samples. Assuming constant dust correction, the best-fit faint-end slope of our LF, $\alpha = -1.62$, is in agreement with these studies.\\

\begin{figure*}
	\epsscale{0.85}
	\plotone{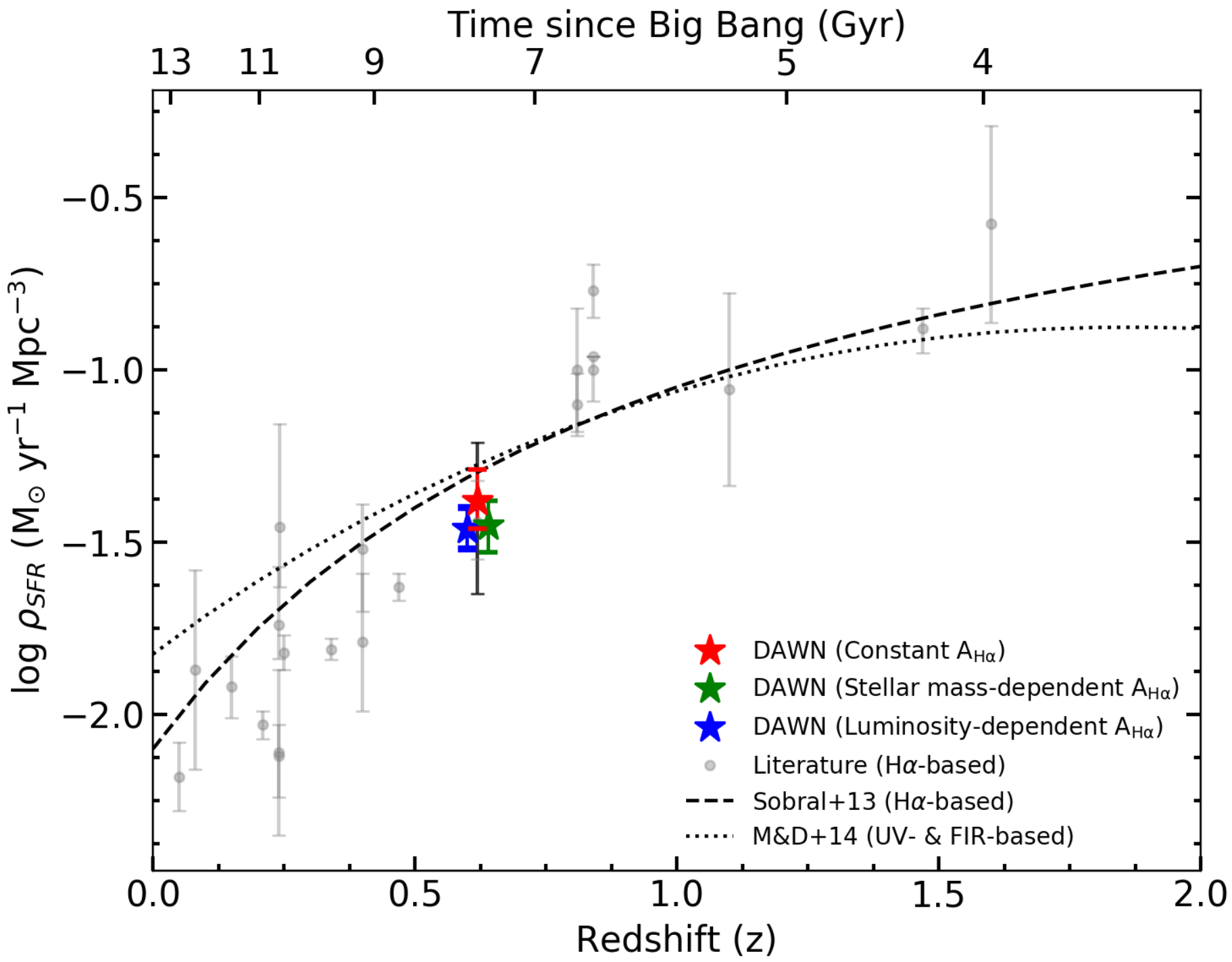}
	\caption{SFRD as a function of redshift. The estimates from DAWN observations (for sake of clarity, different dust-corrected SFRDs have been artificially displaced in redshift) are compared with those from \halpha surveys in the recent past (see Section \ref{sec:sfrd} for more details) and empirical fits for SFRD evolution from \citet{Sobral2013} and \citet{Madau2014}. Within 1$\sigma$ uncertainties, our SFRD estimates are consistent with the observed evolution in SFRD with increasing redshift between $z \sim 0$ and $z \sim 2$ (cosmic variance errors are shown in \emph{black}).}
	\label{fig:sfrd}
\end{figure*}

\subsection{SFR density at $z \sim 0.62$} \label{sec:sfrd}
Using the best-fit Schechter parameters, the total H$\alpha$ luminosity density can be calculated as follows:
\begin{equation}
	\mathcal{L} = L^* \phi^* \Gamma(2+\alpha)
\end{equation}
Following this, the SFR density can be calculated using the standard calibration of \citet{Kennicutt1998}: $\rho_{\mathsf{SFR}} = 7.9 \times 10^{-42}\mathcal{L}$ where $\mathcal{L}$ is calculated by fully integrating down the LFs. 

Although most of the H$\alpha$ luminosity density can be due to active star-formation in galaxies, some contribution is usually attributed to AGN activity as well. Studies in the past have found AGN contamination to be $\sim$10-15\% for \halpha samples at redshifts $z < 2$ \citep[e.g.,][]{Villar2008,Ly2011,Sobral2013}. One way to account for AGN contamination is to look for X-ray identified sources matching with our H$\alpha$ sample. COSMOS2015 catalog contains X-ray sources drawn from XMM-COSMOS \citep{Cappelluti2007,Hasinger2007,Brusa2010} and Chandra-COSMOS \citep{Elvis2009,Civano2012,Civano2016} surveys. The X-ray luminosity limit is $\sim 10^{42}$ erg/s at $z \sim 0.6$ \citep{Marchesi2016} and assuming a typical X-ray to H$\alpha$ ratio, log ($L_X$/$L_{\textnormal{H}\alpha}$) $\sim 1-2$ \citep{Ho2001,Panessa2006,Shi2010}, any AGN-powered H$\alpha$ emitter from either flanking or deep region should be detected in X-rays. We find 5 X-ray matches for our \halpha sample which suggests that AGN contamination is $\sim$2\% of the total sample and $\sim$4\% of the flanking region sub-sample (where it is expected to be complete for X-ray luminous AGN). However, since these X-ray surveys are flux-limited, these matches alone are not a representative of the AGN contamination in our sample.

Another method to assess AGN contamination would be to use the mid-IR color criterion based on the differing spectral energy distributions (SED) of star-forming galaxies and AGN around the rest-frame 1.6$\mu$m bump. In case of AGNs, the SED is a rising power-law after the bump due to the presence of emission from polycyclic aromatic hydrocarbons (PAH) and silicate grains. Using Ks-band and IRAC CH1 (3.6$\mu$m) photometry from COSMOS2015, we measure [$Ks-3.6$] color for our sample where redder colors ($[Ks-3.6] > 0.27$) represent AGNs and bluer colors are mostly star-forming galaxies. Following this criterion, we find 11\% of our sample to be AGN contaminated.

After correcting for AGN contamination, we estimate the SFR density at $z \sim 0.62$ to be, log $\rho_{\mathsf{SFR}} = -1.47$ for luminosity-dependent dust correction, log $\rho_{\mathsf{SFR}} = -1.46$ for stellar mass-dependent dust correction, and log $\rho_{\mathsf{SFR}} = -1.39$ for constant dust correction (Table \ref{tab:LF+SFRD}). For DAWN-COSMOS, the cosmic variance uncertainties were estimated to be around 23\% based on calculations from \citet{DR2010}. Figure \ref{fig:sfrd} shows a comparison of SFR density as a function of redshift for DAWN and other H$\alpha$ based surveys (e.g., \citealt{Ly2007,Morioka2008,Shioya2008,Villar2008,Sobral2009,Westra2010,Ly2011,Sobral2013,Stroe2015,GG2016,Khostovan2020}. We also compare our SFR density estimate to the empirical fits given by \cite{Sobral2013} and \cite{Madau2014}. Using H$\alpha$ samples at $z\sim0.4, 0.8, 1.47, 2.23$ from HiZELS, \cite{Sobral2013} provide an empirical fit for SFR density as a function of redshift, log $\rho_{\mathsf{SFR}} = -2.1/(z+1)$. In \cite{Madau2014}, an empirical fit for SFR density is derived based on measurements from a host of recent UV and IR galaxy surveys. Within 1$\sigma$ uncertainties, our SFR density estimates are consistent with these fits as shown in Figure \ref{fig:sfrd}. $\rho_{\mathsf{SFR}}$ based on luminosity-dependent dust correction is slightly lower mostly due to the fact that the correction is unequal across the sample given the luminosity-dependent relation and and also that some of the faintest \halpha emitters in the sample require no correction, according to this dust correction method.

\section{Summary and Conclusions} \label{sec:concl}
In this work, we have presented new measurements of the H$\alpha$ LF and SFR density for NB-selected galaxies at $z \sim 0.62$ from the DAWN survey. Compared to \citetalias{Coughlin2018}, an additional area of 1.23 deg$^2$ was surveyed in the COSMOS region with a resulting total area coverage of $\sim$1.5 deg$^2$ and co-moving volume of $\sim3.5\times10^4$ Mpc$^3$ at $z \sim 0.62$. In the deepest COSMOS region, the survey reaches a 5$\sigma$ emission-line flux depth of $\sim7.7\times10^{-18}$ erg s$^{-1}$ cm$^{-2}$. The main findings of this work are as follows:
\begin{enumerate}[(1)]
	\item A total of 1,163 sources were selected as NB-excess emitters with EW$_{\text{obs}} \geqslant 18$\AA\ and color-significance $\geqslant3\sigma$. Among them, 241 were classified as H$\alpha$ emitters at $z\sim0.62$ based on a combination of spectrophotometric and color-color criteria with up to 111 confirmations from previous spectroscopic surveys. 
	
	\item H$\alpha$ LFs were constructed after accounting for [N\textsc{ii}] contamination, completeness correction, and dust attenuation. Given the ambiguity surrounding different methods of dust correction, the three most popular methods were used: the luminosity-dependent correction following \cite{Hopkins2001}, stellar mass-dependent correction following \cite{GB2010} and a constant correction of A$_{\text{H}\alpha}=1$. All three LFs are well described by a Schechter function with best-fit values of L$^*$ = $10^{42.31}$ erg s$^{-1}$, $\phi^*$ = $10^{-2.8}$ Mpc$^{-3}$, $\alpha = -1.39$ (luminosity-dependent dust correction), L$^*$ = $10^{42.36}$ erg s$^{-1}$, $\phi^*$ = $10^{-2.91}$ Mpc$^{-3}$, $\alpha = -1.48$ (stellar mass-dependent dust correction), and L$^*$ = $10^{42.24}$ erg s$^{-1}$, $\phi^*$ =$10^{-2.85}$ Mpc$^{-3}$, $\alpha = -1.62$ (constant dust correction). At $z\sim0.62$, the LFs as well as the Schechter parameters are in good agreement with the expected evolution in comparison to those at other redshifts ($0 < z < 2$). Within the 1$\sigma$ uncertainties, the Schechter parameters are also in good agreement with the empirical relation given by \cite{Sobral2013}. However, the derived faint-end slope is shallowest for luminosity-dependent dust correction ($\alpha=-1.39$), and steepest for constant dust correction ($\alpha=-1.62$).
	
	\item On fully integrating the H$\alpha$ LF, we obtain a total H$\alpha$ luminosity density of $\mathcal{L} = 10^{39.68}$ erg s$^{-1}$ Mpc$^{-3}$, in case of luminosity-dependent dust correction, $\mathcal{L} = 10^{39.69}$ erg s$^{-1}$ Mpc$^{-3}$ for stellar mass-dependent dust correction and $\mathcal{L} = 10^{39.76}$ erg s$^{-1}$ Mpc$^{-3}$ for constant dust correction. Following the standard calibration from \cite{Kennicutt1998}, the SFR density at $z\sim0.62$ is estimated to be, $\rho_{\mathsf{SFR}} = 10^{-1.47}$ M$_{\odot}$yr$^{-1}$Mpc$^{-3}$ for luminosity-dependent dust correction,$\rho_{\mathsf{SFR}} = 10^{-1.46}$ M$_{\odot}$yr$^{-1}$Mpc$^{-3}$ for stellar mass-dependent dust correction, and $\rho_{\mathsf{SFR}} = 10^{-1.39}$ M$_{\odot}$yr$^{-1}$Mpc$^{-3}$ for constant dust correction, which are highly consistent with the evolution of SFR densities across the redshift range, $0 < z < 2$, as seen from previous \halpha surveys.
\end{enumerate}

Among H$\alpha$ studies at low redshifts ($0<z<1$), this survey fills the gap that exists at $z\sim0.62$, and it is the only survey to comprehensively study both the faint and bright end of the LF at this redshift. Moreover, this work illustrates the importance of combining observations that are significantly deep (compared to L$^*$) with observations covering a substantial volume (compared to $1/\phi^*$), in order to better constrain the entire luminosity function.

\acknowledgments
We thank NOAO for the generous allocation of observing time for the DAWN survey, and the Kitt Peak staff for their expert support of DAWN observations. We thank Kimberly Emig, Raviteja Nallapu, Emily Neel, Mark Smith, Stephanie Stawinski, Jacob Trahan, Nicholas Valverde, Trevor Van Engelhoven, Sherman Florez, Amy Robertson, Cristian Soto, Jonathan Florez, Dave Bell, Sofia Rajas and Karen Butler for their help with the DAWN survey observing runs. We thank the US National Science Foundation for its financial support through NSF grant AST-0808165, which supported our custom narrow-band filter purchase, and AST-1518057, which supported our data analysis. We also thank NASA for financial support via WFIRST Preparatory Science Grant NNX15AJ79G and WFIRST Science Investigation Team contract NNG16PJ33C, which provided additional support for our scientific analysis. SV acknowledges support from a Raymond and Beverley Sackler Distinguished Visitor Fellowship and thanks the host institute, the Institute of Astronomy, where this work was concluded. SV also acknowledges support by the Science and Technology Facilities Council (STFC) and by the Kavli Institute for Cosmology, Cambridge. JXW thanks support from NSFC 11421303 and 11890693. This work is based on data products from observations made with ESO Telescopes at the La Silla Paranal Observatory under ESO programme ID 179.A-2005 and on data products produced by TERAPIX and the Cambridge Astronomy Survey Unit on behalf of the UltraVISTA consortium. This work has made use of the following open-source softwares: SciPy, NumPy, Matplotlib \citep{Jones2001,Oliphant2006,Hunter2007}, Astropy \citep{Robitaille2013}, corner.py \citep{FM2016}, DS9 \citep{Joye2003} and TOPCAT \citep{Taylor2005}.

\bibliographystyle{aasjournal}
\bibliography{ref}



\end{document}